\documentclass[12pt,twoside]{article}      
  \usepackage{amsmath, amsfonts,pst-feyn,pst-key}

\newfont{\banner}{cmssdc10  scaled 1820}
\newfont{\figuretitle}{cmssdc10  scaled 1440}
\title{Applications of Chern-Simons  Ward Identities to 
$ \log(T\tau) $ Conductivity 
Calculations of the $\nu=1/2$ System } 
\author{J\"urgen Dietel \\ 
Institut f\"ur Theoretische Physik, \\ Universit\"at Leipzig, 
Germany }      

\date{}      

\begin{document}             
\psset{ArrowInside=->, ArrowAdjust=true}

\maketitle 
\begin{abstract}
We reconsider the theory of the half-filled Landau level with impurities 
using the Chern-Simons formulation and study Ward identities for 
the Chern-Simons
theory. From these we get conductivity diagrams with impurities 
which obey the continuity equation. We calculate the conductivity of these 
diagrams for which we obtain $ \log(T\tau) $ divergent 
conductivity diagrams for 
low temperature $ T $. We compare our result with the experimental values
of the low temperature conductivities. Finally we calculate the conductivity 
for small deviations of the magnetic field $ B $ from the value $\nu=1/2 $.
In this case we get a singularity in the conductivity.       
\end{abstract}

\section{Introduction}
In this paper we consider a system of electrons in a strong magnetic 
field in two dimensions. This system is characterized by 
the filling factor $ \nu $, defined as the electron density divided by the 
density of a completely 
filled Landau level. In the case of $ \nu \approx 1/2 $ the 
behavior of the system resembles that of a Fermi liquid in the 
absence of a magnet field or at small magnetic fields. Over past years an intriguing picture has emerged: 
At $ \nu=1/2 $ each electron combines with two flux quanta of the magnetic field to form a composite fermion (CF); these composite fermions 
then move in an effective
magnetic field which is zero on the average. The interpretation of 
many experiments support this picture. We mention some transport experiments
for illustration \cite{wi1}.\\
The theoretical framework for the understanding of the $ \nu=1/2 $-system 
was 
developed by Halperin, Lee and Read \cite{hlr}. In their theory one has 
to pay a 
price to get a mean-field free system for the CFs. The CFs 
interact via long-ranged gauge interactions. Further in real $ GaAs/Al_xGa_{1-x}As$ 
heterojunctions one also has a large number of Coulomb impurities which  
has to be integrated in the theory, too. For the case of the Coulomb problem 
this was done about 20 years ago \cite{al1, al2}. \\
Our aim is to calculate the low temperature conductivity of CFs 
with impurities. 
In the first section of this work we care about which Feynman graphs should be 
calculated to get a good conductivity  for the $ \nu=1/2 $ system. This 
question will be treated with the help of Ward identities 
\cite{ab1} of Chern-Simons systems. In the second section we will 
apply the results of section 1 to get some Feynman-graphs which should
give a good approximation of the conductivity for CFs. At least  
 we will calculate the conductivity of these Feynman graphs 
at low temperature $ T $ and make comparison with experimental results.  
 
\section{The Impact of Ward Identities of CFs on the 
perturbational continuity equation}

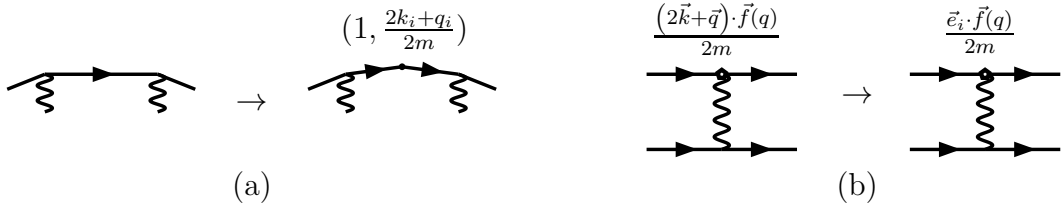
\begin{figure}[t]
 {
  \psset{unit=1cm}
\begin{pspicture}(10,3)
\psset{linewidth=1.5pt,arrowinset=0}
\psline[linewidth=1.3pt,ArrowInside=->](0.5,1.7)(2.0,1.7)
\psline[linewidth=1.3pt,ArrowInside=-](0.,1.5)(0.5,1.7)
\psline[linewidth=1.3pt,ArrowInside=-](2.0,1.7)(2.5,1.5)
\WavyLine[n=25,beta=180](0.5,1.7)(0.5,1.2)
\WavyLine[n=25,beta=180](2.0,1.7)(2.0,1.2)
\rput(3.25,1.3){$\rightarrow$}

\psline[linewidth=1.3pt,ArrowInside=->](4.5,1.7)(5.25,1.8)
\psline[linewidth=1.3pt,ArrowInside=->](5.25,1.8)(6.,1.7)
\rput(5.35,2.3) {$(1,\frac{2k_i+q_i}{2m})$ }
\psline[linewidth=1.3pt,ArrowInside=-](4.,1.5)(4.5,1.7)
\psline[linewidth=1.3pt,ArrowInside=-](6.,1.7)(6.5,1.5)
\WavyLine[n=25,beta=180](4.5,1.7)(4.5,1.2)
\WavyLine[n=25,beta=180](6.,1.7)(6.,1.2)
\psdots[dotstyle=*,dotscale=0.6](5.25,1.8)
\rput(3.25,0.2) { (a) } 

\psline[linewidth=1.3pt,ArrowInside=->](8.5,1.7)(9.5,1.7)
\psline[linewidth=1.3pt,ArrowInside=->](9.5,1.7)(10.5,1.7)
\WavyLine[n=25,beta=180](9.5,1.7)(9.5,0.7)
\psline[linewidth=1.3pt,ArrowInside=->](8.5,0.7)(9.5,0.7)
\psline[linewidth=1.3pt,ArrowInside=->](9.5,0.7)(10.5,0.7)
\psdots[dotstyle=pentagon,dotscale=1.1](9.5,1.7)
\rput(11.3,1.4){$\rightarrow$}
\rput(9.5,2.3) {$\frac{\left(2\vec{k}+\vec{q}\right)\cdot\vec{f}(q)}{2m}$ }
\psline[linewidth=1.3pt,ArrowInside=->](12,1.7)(13,1.7)
\psline[linewidth=1.3pt,ArrowInside=->](13,1.7)(14,1.7)
\WavyLine[n=25,beta=180](13,1.7)(13,0.7)
\psline[linewidth=1.3pt,ArrowInside=->](12,0.7)(13,0.7)
\psline[linewidth=1.3pt,ArrowInside=->](13,0.7)(14,0.7)
\psdots[dotstyle=pentagon,dotscale=1.1](13,1.7)
\rput(11.3,0.2) { (b) } 
\rput(13,2.3) {$\frac{\vec{e}_i\cdot\vec{f}(q)}{2m}$ }
\end{pspicture}
 \caption{Vertexoperations for constructing $ \Lambda^b $ from a 
 self energy graph $ \Sigma^b $. For $ \Lambda^b_0 $ one has to do 
 vertexoperation (a) with a 
 density-vertex between the two Green's functions. 
 For $ \Lambda^b_i $ one has to do vertex operation (a) with a 
 current-vertex between the two Green's functions as well as the 
 vertex operation (b).}
}
\end{figure}

In this section we try to answer the question which subset of 
Feynman diagrams should be calculated, in order to get a good estimate for  
the conductivity of the $ \nu=1/2 $-
system.  
To answer this question we consider at first some 
self energy diagrams $ \Sigma^p(k) $ of 
frequency $ k_0 $ and impulse $(k_1,k_2)$  (Index $ p $ for perturbational). 
In the following we will construct $ T $-product response Feynman-graphs 
from $ \Sigma^p $. 
$ \Lambda_\mu $ ( $ \mu=0..2 $ ) is defined as 
\begin{equation} \label{eq1}
\left\langle\left|T\left[j_\mu(q),\Psi^*(k)\Psi(k+q)\right]\right|\right\rangle
= G(k) \Lambda_\mu(k+q,q) G(k+q)\;, 
\end{equation}
where $ G(k) $ is the Chern-Simons Green's function.
 $ \vec{j} $, $ \Psi $ are the current operator and field operator  
of the Chern-Simons theory. 
In the following one has to notice that the Chern-Simons vertices 
contain current vertices   
$ \frac{(2\vec{k}+\vec{q})\vec{f}(q)}{2m} $ where 
$ f_i(q)=i 2 \pi \tilde{\phi}\epsilon_{ij} \frac{q_j}{q^2}$.   
We now apply some transformations on $ \Sigma^p $ to get 
$ \Lambda_\mu $ graphs. 
We use the notation $ \Lambda^p_0 $ for 
$ \Lambda_0 $ diagrams which originate from   
$ \Sigma^p $ through the operation (a) in figure 1 for every Green's 
function in $ \Sigma^p $ .
 This means that one  
one has to 'divide' every Green's function in $ \Sigma^p $ making two 
Green's function. Similarly  $ \Lambda^p_i$ ($ i=1,2 $) denotes the 
diagrams which are originating through operation (a), (b) in figure 1.
This has to be done for every 
Green's function and current vertex in $ \Sigma^p$.
In contrast to $ \Lambda^p_0 $ operation (a) puts for $ \Lambda^p_i$  
a current operator between the two Green's functions. 
The vertices in diagram (b) are current CS-vertices. In (b)   
every current vertex of $ \Sigma^b $ is transformed into a density-density 
CS current coupling.   
Then one can derive the following Ward identities   
\begin{eqnarray} 
-q_0 \left(\Lambda^p_0(k,k+q_0e_0)-1\right) & = & \Sigma^p(k+q_0e_0)
-\Sigma^p(k) \label{eq2}\; , \\  
q_i \left(\Lambda^p_{i}(k,k+q_ie_i)-q_i\frac{2 k_{i}+q_i}{2 m}\right) & = & 
\Sigma^p(k+q_ie_i)-\Sigma^p(k)\;. \label{eq3}
\end{eqnarray} 
To prove these relations we classify the Green's functions $ G^0_i $
(the free Green's functions)  
of the self energy graph $ \Sigma^p $. $ G^0_{i_1} \sim G^0_{i_2} \in W_j \in {\cal W} $
if there is a path in the graph $ \Sigma^p $ which do not pass a vertex, 
connecting $ G^0_{i_1} $ with $ G^0_{i_2} $. 
$ G^0_a $, $G^0_e \in W_{ae} $ denote the Green's functions which are at 
the continuation of the outer truncated Green's functions
of the graph $\Sigma^p $. In the following we show at first 
relation (\ref{eq2}).
$ \Sigma(k+q_0e_0) $ is the graph in which one makes the substitution 
$ G^0_i (k') \rightarrow G^0_i(k'+q_0 e_0) $ for every Green's function in 
$\Sigma^p$.
Furthermore we denote as $ \Gamma^p_i(k+q,k;k',k'+q) $ 
the graph in which   
a Green's function $ G^0_i $ is eleminated from $ \Sigma^p $. 
With the help of $ \{G^0_{1},G^0_{2}.. \}=W_j \not= W_{ae}$ one gets 
\begin{equation} \label{eq4}
\sum\limits_{G^0_i \in W_j} \sum\limits_{k'} 
\Gamma_{i}(k,k+q_0e_0;k'+qe_0,k')\left(
G^0_{i}(k'+q_0e_0)-G^0_{i}(k')\right)\quad = \quad 0 \;.
\end{equation}
This is valid because the Green's functions in $ W_j \not= W_{ae}$
are forming a circle.
Furthermore one gets for $ \{G^0_{1},G^0_{2}.. \}= W_{ae}$
\begin{eqnarray} 
\lefteqn{\sum\limits_{G^0_i \in W_{ae}} \sum\limits_{k'} 
\Gamma^p_{i}(k,k+q_0e_0;k'+q_0e_0,k')
\left(G^0_{i}(k'+q_0e_0)-G^0_{i}(k')\right)}
\label{eq5}\\ 
& = & \sum\limits_{k'}
 \Gamma^p_{e}(k,k+q_0e_0;k'+q_0e_0,k')G^0_{e}(k'+q_0e_0)  
-\Gamma^p_{a}(k,k+q_0e_0;k'+q_0e_0,k')G^0_{a}(k') \nonumber \\
 & = & \Sigma^p(k+q_0 e_0)-\Sigma^p(k) \;. \nonumber 
\end{eqnarray}
With the help of (\ref{eq4}), (\ref{eq5}), one gets (\ref{eq2}). \\
We now prove the current Ward identity (\ref{eq3}). 
If one has no current vertex in $ \Sigma^p $ one gets immediately 
(\ref{eq3}) in the same manner as (\ref{eq2}). Otherwise 
we denote $ C $ as the number of current vertices.
Furthermore we denote $ B(G^0_i) $ ($ E(G^0(i)$) as the beginning- (end-)
vertex of the directed Green's function $ G^0_i $. 
If $ B(G^0_i) \in C $ ($E(G^0_i)\in C$), we denote  
$ \vec{j}(B(G^0_i)) $ ($\vec{j}(E(G^0_i)$)  as the current of the vertex 
$ B(G^0_i)$  ($ E(G^0_i) $).  
Relations (\ref{eq4}), (\ref{eq5}) keep their validity if one makes 
the following substitutions: Every Green's function $ G^0_i(k'+q_0 e_0) $
($ G^0_i(k'+q_0 e_0) $)
in the sums of (\ref{eq4}), (\ref{eq5})
should be replaced by $ G^0_i(k'+q_i e_i)$ ($ G^0_i(k')$) if  
$ E(G^0_i) \notin S $ ($B(G^0_i)\notin S$), 
otherwise by 
$ G^0_i(k'+q_i e_i)\left( 1+ 
\left[\vec{j}(E(G^0_i)) \rightarrow \frac{q_i}{2m} e_i \right]\right) $  
$ \left(G^0_i(k')\left( 1- 
\left[\vec{j}(B(G^0_i)) \rightarrow \frac{q_i}{2m} e_i \right]\right) \right)$.
The brackets have the meaning of 
replacing the 
current $ \vec{j} $ of the vertex $ B(G_i) $ ($E(G_i)$) in 
$ \Gamma^p_i $ by $ \frac{q_i}{2m} e_i $. With the help of 
these substitutions in the expressions (\ref{eq4}), (\ref{eq5}) one immediately gets the 
current Ward identity (\ref{eq3}).\\
\vspace{0.5cm}
Similar to (\ref{eq2}), (\ref{eq3}) one can show with $ q=(q_0,q_1,q_2) $ 
the following combined Ward identity 
\begin{equation} \label{eq8}
-q_0\Lambda^p_0(k,k+q)+\sum\limits_{i=1}^2 q_i\Lambda^p_{i}(k,k+qe_i)
+q_0-\sum\limits_{i=1}^2 q_i\frac{2 k_i+q_i}{2 m}= \Sigma^p(k+q)-\Sigma^p(k)
\;. 
\end{equation}     
We now consider our primary problem. It is clear that the 
response (density or current) to an external 
$ \vec{A} $-field should obey 
the continuity equation. We now define the approximation of the 
density-current and 
current-current $ T $-product as 
\begin{eqnarray} \label{eq9}
\lefteqn{\hspace{-2cm}
\left\langle\left|j_\mu(q)j_\nu(-q)\right|\right\rangle \approx 
\Pi^p_{\mu i}(q)
=  \int \sum\limits_{k'} G^{\Sigma}(k'+q) \Lambda ^p_\mu(k'+q,k')
G^{\Sigma}(k')}  \\
& & \hspace{3cm}
\times \left(\delta_{\mu,0}+(1-\delta_{\mu,0})\frac{2
    k_\mu'+q_\mu}{2m}\right) \;,\nonumber
\end{eqnarray}
where $ G^\Sigma(k)=(k_0-\xi(\vec{k})-\Sigma^p(k))^{-1} $ \\
Within the help of the combined Ward identities (\ref{eq8}), 
the Kubo-formula and   
the relation of $ T $-products and retarded correlation functions 
\cite{ab1},
one sees that the continuity equation is valid 
if $ \Pi^p_{\mu \nu} $ fulfills the equation
\begin{equation} \label{eq10}
 e^2 q_0 \Pi^p_{0 i}(q)-e^2 \sum\limits_{j=1}^2 q_j \Pi^p_{j i}(q)=-i q_i 
\frac{e^2 n_e}{m}\;,
\end{equation}
where $ n_e $ is the electron density. \\
This equation can easily checked with the help of (\ref{eq8}).
Now we have the problem that the graphs $\Pi^p_{\mu \nu}(q) $ are not 
symmetric in the coupling of the currents. 
On one side of $\Pi^p_{\mu \nu}(q) $ one
has density-density Chern-Simons current-couplings. 
These current couplings are missing 
on the other side of the graphs $\Pi^p_{\mu \nu}(q) $. To get rid of 
this problem we name $ \Lambda^{b,4}_{i} $ as the members of
$ \Lambda^{b}_i $ which are originating through operation (b) in figure 1 from 
$ \Sigma^b $. We now close the open ends of $\Lambda^{b,4}_{i}  $ with a 
Green's function $ G^0 $and call these graphs ${\Lambda}^{b,4}_{i}  $. 
These  graphs consists of closed Green's function loops.
To these graphs we apply the operations (a) or (b) of figure 1, respectively, 
and call these graphs $ \Xi_{0 i}(q)$ or $ \Xi_{j i}(q) $, respectively .
Than one sees similar to the proof of (\ref{eq2}), (\ref{eq3}) 
that $ \Xi_{0 i}(q) $, $ \Xi_{j i}(q) $ fulfills the equation 
$ e^2 q_0 \Xi^p_{0 i}(q)-e^2 \sum\limits_{j=1}^2 q_j \Xi^p_{j i}(q)=0 $
($ \Xi_{0 i}(q) $, $ \Xi_{j i}(q) $ connsists only of closed Green's function  
circles). 
With the definition $ {\Pi}^{b,2}_{\mu \nu}=\Pi^{b}_{\mu \nu}+\Xi^p_{\mu \nu} $one sees that $ {\Pi}^{b,2}$ consists of current symmetric graphs. 
Furthermore $ {\Pi}^{b,2}$  fulfills equation (\ref{eq10}).   \\
It is also clear that the continity equation is fulfilled 
if one only considers the graphs in $ {\Pi}^{b,2}$ which have the same number 
of vertices. 

\section{The Calculation of the Conductivity}    

Next we calculate the conductivity of a $ \nu=1/2 $  
Chern-Simons gas with impurities. The calculable quantity is the conductivity 
of CFs which is related to the physical conductivity through 
$ \sigma^{CS}=\sigma^2_{xx}+\sigma_{xy}/\sigma_{xx} $ \cite{hlr}.
For the mean-field Green's function of CFs in an impurity 
background one has 
$ G_{\pm}(\omega,p)=
\left(\omega-\xi(p) \pm i/(2\tau)\right)^{-1} $.     
In the following calculation  
we need momentum integrals $ \int\limits_{0}^{\infty} dp $ of products of 
such Green's functions. One can approximate
these momentum integrals (also for $ k_F l\approx 1 $) by extending the range 
of such integrals to infinity, $ \int\limits_{-\infty}^{\infty} dp $. 
We use this  approximation in the following.   
In doing that we hope to get better results for smaller $ k_F l $
in contrast to the standard $ k_F l \gg 1 $ approximation. 
To calculate physical 
quantities also for smaller densities of the $ \nu=1/2 $-system is suggestive 
because one could reduce the effective $ k_F l $ of a $\nu=1/2$-system 
in increasing 
the magnetic field to get a $ \nu=(5/2,9/2,13/2 ..)$-system. Systems with these fillings behave similar to $ \nu=1/2 $-systems with a reduced density 
$ n_e=(n_e/5, n_e/9,n_e/13..) $.\\
In the following we limit our calculation to the case of 
impuritis with a $ \delta $-corrolation \cite{hlr}.  
It could be  shown \cite{fa1,kh1} that one gets the same 
results in the case of RMF-scattering for $ k_Fl\gg 1 $.
Further we will only discuss the particle-hole channel conductivity graphs 
because we don't expect any weak localization correction due to the magnetic 
broken time reversal symmetry \cite{hlr}. \\         
In the Coulomb gauge ($ div \vec{A}=0 $), the gauge interaction between 
CFs at $ \omega \tau \ll 1 $ and $ql \ll 1 $ is  
described by  
\begin{equation} \label{eq11}
\begin{array}{c c c}
D^{-1}_{\mu \nu} & = & \left(
\begin{array} {c c}
\frac{m}{2\pi} \frac{Dq^2}{Dq^2-i \omega} & -i \frac{q}{4 \pi} \\
i \frac{q}{4 \pi}  & -i \gamma_q |\omega|+\chi_q q^2
\end{array}
\right) \;, 
\end{array}
\end{equation}
where $ D=\frac{1}{2} v_F^2 \tau $ is a diffusion coefficient, 
$ \chi_q=\frac{1}{24 \pi m} +\frac{V_q}{(4 \pi)^2} $ is an effective 
diamagnetic susceptibility given in terms of the 
pairwise electron potential $ V_q $, 
$ \gamma_q=\frac{mD}{2\pi} $  is proportional
 to the CF mean free path $ l= v_F \tau $ and $m$ stands for the 
effective mass.
Comparing the gauge interaction functions (\ref{eq11}) with the gauge 
interaction functions in \cite{hlr,kh1} one sees that we get a difference 
by a factor $ 1/2 $ in $ \chi_q $. 
In \cite{di1} we made an exakt calculation of the  polarisator 
$ \Pi_{11} $ without impurities. In this polarisator one also gets
half the value of the magnetic susceptibility in \cite{hlr}. 
So we believe that our calculation is correct.
As in \cite{kh1} we will discuss in the following the two 
regimes  $ V_q \approx V_0=2 \pi \frac{e^2}{\kappa} $ and 
$ V_q = 2\pi \frac{e^2}{q} $.   \\
We now generate the conductivity diagrams from  
the left hand side self energy diagrams in figure 2. These diagrams are also 
the starting point in the case of the Coulomb gas \cite{al1,al2}. Then one 
gets
among others the diagrams in the right hand side of figure 2. 
In \cite{al2} was 
shown that the conductivity of the diagrams not listed 
in figure 2 adds to zero. This can also be shown for our 
$ k_F l\approx 1 $ approximation. This is also significant for CFs. 
We now invert the matrix (\ref{eq11}). So we get     
\begin{eqnarray} 
D_{00} & = & \frac{-i|\omega|+Dq^2}{\frac{m}{2\pi}D q^2-\frac{q^2}{16\pi^2} 
 \frac{-i\, |\omega|+Dq^2}{-i \gamma_q |\omega|+\chi_q q^2}}\;, \label{eq12}\\
 D_{11} & = & \frac{1}{-i\gamma_q |\omega|+\chi_q q^2-\frac{q^2}{16\pi^2}   
 \frac{-i\,|\omega|+Dq^2}{\frac{m}{2\pi}D q^2}}  \approx
 \frac{1}{-i \gamma_q |\omega|+\chi'_q q^2} \,\theta(Dq^2-|\omega|) \;,
\label{eq13}
\end{eqnarray}
where $ \chi'_q=\chi_q-\frac{1}{8\pi m} $.


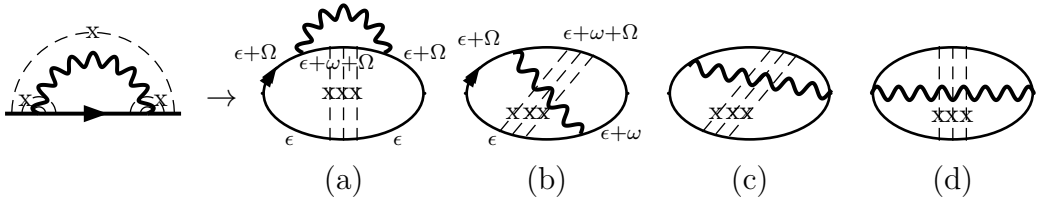
\begin{figure}[t]
 {
  \psset{unit=0.9cm}
\begin{pspicture}[-0.2](10,2)
\psset{ArrowInside=->, ArrowAdjust=true}
\psset{linewidth=1.5pt,arrowinset=0}
\WavyArc[linewidth=1.5pt,n=20,beta=180](1.3,0.7){0.8}{0}{180}
\psline[ArrowInside=->](0.5,0.7)(2.1,0.7)
\psline[ArrowInside=-](0.,0.7)(0.5,0.7)
\psline[ArrowInside=-](2.1,0.7)(2.6,0.7)
\psset{linewidth=0.5pt,arrowinset=0}
\psarcn[linewidth=0.5pt,linestyle=dashed]{-}(0.5,0.7){0.1}{180}{0}
\psarcn[linewidth=0.5pt,linestyle=dashed]{-}(0.5,0.7){0.25}{180}{0}
\psarcn[linewidth=0.5pt,linestyle=dashed]{-}(2.1,0.7){0.1}{180}{0}
\psarcn[linewidth=0.5pt,linestyle=dashed]{-}(2.1,0.7){0.25}{180}{0}
\psarcn[linewidth=0.5pt,linestyle=dashed]{-}(1.3,0.7){1.2}{180}{0}
\rput(0.3,0.9){\footnotesize x}
\rput(2.3,0.9){\footnotesize x}
\rput(1.3,1.9){\footnotesize x}
\rput(3.2,0.9){$\rightarrow$}


\psset{linewidth=1pt,arrowinset=0}
\psellipse(5,1)(1.2,0.7)
\WavyArc[linewidth=1.5pt,n=20,beta=180](5,1.6){0.6}{0}{180}
\psline[linewidth=0.5pt,linestyle=dashed,ArrowInside=-](5,0.3)(5,1.7)
\psline[linewidth=0.5pt,linestyle=dashed,ArrowInside=-](4.8,0.3)(4.8,1.7)
\psline[linewidth=0.5pt,linestyle=dashed,ArrowInside=-](5.2,0.3)(5.2,1.7)
\rput(5.05,1){\footnotesize x }
\rput(4.85,1){\footnotesize x }
\rput(5.25,1){\footnotesize x }
\psdots[dotstyle=*,dotscale=0.4](3.8,1)
\psdots[dotstyle=*,dotscale=0.4](6.2,1)
\rput(5,-0.3) { (a) } 
\rput(3.7,1.6) { $ \scriptstyle \epsilon+\Omega $ }
\rput(6.2,1.6) { $ \scriptstyle \epsilon+\Omega $ }
\rput(4.2,0.3) { $ \scriptstyle \epsilon$ }
\rput(5.8,0.3) { $ \scriptstyle \epsilon$ }
\rput(4.9,1.45) { $ \scriptstyle \epsilon+\omega+\Omega $ }
\psline[linewidth=0.9pt] {->}(3.88,1.2)(3.98,1.35)
\rput(7.5,0.7){\footnotesize x}
\rput(7.95,0.7){\footnotesize x}
\rput(7.75,0.7){\footnotesize x}
\psellipse(8,1)(1.2,0.7)
\WavyLine[linewidth=1.5pt,n=20,beta=180](7.5,1.6)(8.5,0.4)
\psline[linewidth=0.5pt,linestyle=dashed,ArrowInside=-](7.5,0.4)(8.5,1.6)
\psline[linewidth=0.5pt,linestyle=dashed,ArrowInside=-](7.3,0.43)(8.3,1.63)
\psline[linewidth=0.5pt,linestyle=dashed,ArrowInside=-](7.7,0.37)(8.7,1.57)
\psdots[dotstyle=*,dotscale=0.4](6.8,1)
\psdots[dotstyle=*,dotscale=0.4](9.2,1)
\rput(8,-0.3) { (b) } 
\rput(7.0,1.8) { $ \scriptstyle \epsilon+\Omega $ }
\rput(8.8,1.9) { $ \scriptstyle \epsilon+\omega+\Omega $ }
\rput(7.2,0.3) { $ \scriptstyle \epsilon$ }
\rput(9.1,0.4) { $ \scriptstyle \epsilon+\omega$ }
\psline[linewidth=0.9pt] {->}(6.88,1.2)(6.98,1.35)
\psellipse(11,1)(1.2,0.7)
\rput(10.5,0.7){\footnotesize x}
\rput(10.95,0.7){\footnotesize x}
\rput(10.75,0.7){\footnotesize x}
\WavyLine[linewidth=1.5pt,n=20,beta=30](10.1,1.4)(12.2,1)
\psline[linewidth=0.5pt,linestyle=dashed,ArrowInside=-](10.5,0.4)(11.5,1.6)
\psline[linewidth=0.5pt,linestyle=dashed,ArrowInside=-](10.3,0.43)(11.3,1.63)
\psline[linewidth=0.5pt,linestyle=dashed,ArrowInside=-](10.7,0.37)(11.7,1.57)
\psdots[dotstyle=*,dotscale=0.4](9.8,1)
\psdots[dotstyle=*,dotscale=0.4](12.2,1)
\rput(11,-0.3) { (c) } 
%

\psellipse(14,1)(1.2,0.7)
\WavyLine[linewidth=1.5pt,n=20,beta=180](12.8,1)(15.2,1)
\psline[linewidth=0.5pt,linestyle=dashed,ArrowInside=-](14,0.3)(14,1.7)
\psline[linewidth=0.5pt,linestyle=dashed,ArrowInside=-](13.8,0.3)(13.8,1.7)
\psline[linewidth=0.5pt,linestyle=dashed,ArrowInside=-](14.2,0.3)(14.2,1.7)
\rput(14.05,0.65){\footnotesize x }
\rput(13.85,0.65){\footnotesize x }
\rput(14.265,0.65){\footnotesize x }
\psdots[dotstyle=*,dotscale=0.4](12.8,1)
\psdots[dotstyle=*,dotscale=0.4](15.2,1)
\rput(14,-0.3) { (d) } 
\end{pspicture}
 \caption{The relevant conductivity diagrams discussed in the text. 
On the left hand side one sees the self energy diagram from which the 
conductivity diagrams are extracted through the formalism of section 2.}
}
\end{figure} 
One can further see from (\ref{eq11}) that the vertex $ D_{01} $ isn't 
as much divergent as $ D_{00} $, $D_{11} $. 
Because of this one can show, that this vertex gives no 
$ \log(T \tau) $-term in the conductivity.   
We now calculate the graphs (a), (b) of figure 2.
At first we will calculate the graphs (a), (b) with the $ D_{00} $-vertex. 
To get the divergent part of the graphs one has to put the vertex correction 
$ \Gamma(\epsilon,\omega,q)=(m \tau)^{-1}(D q^2-i |\omega|)^{-1} $ 
for $ \epsilon (\epsilon+\omega) <0 $ 
 at the endpoints of the vertex $ D_{00} $. One can then 
use the analysis of the Coulomb problem \cite{al2}.
After integration one gets for $ D \gamma_q \gg \chi_q $ 
\begin{equation} \label{eq14}
\delta \sigma^{D_{00}}_{ii}\approx\frac{1}{\pi h} \left(\frac{(k_F l)^2}
{(\frac{5}{3}+(k_F l)^2)}+ \frac{2}{3} \frac{(k_F l)^2}{(\frac{5}{3}+(k_F l)^2)^2}
\log\left(\frac{3}{2}(k_F l)^2\right)\right)\log(T \tau) \;,
 \end{equation}
where $ l=(h k_F \tau)/(2\pi m) $.\\ 
From (\ref{eq14}) one sees that in the case $ (k_F l) \gg 1 $, 
$ \sigma_{ii}^{D_0} $ goes to the conductivity of the Coulomb problem \cite{al1}.
We now calculate the conductivity of the $ D_{11} $-vertex. 
To calculate the diagrams (a), (b) of figure 2 one needs the quantities 
$ J^\pm_i(\vec{q})=
\sum_{\vec{p}} G_{\pm}^2(p) G_{\mp}(p)\frac{\vec{p}_i}{m}(
\frac{\vec{p}}{m} 
\times \frac{\vec{q}}{q})=(\mp i \epsilon_F \tau^2-\tau/2) 
(\vec{e}_i \times \frac{\vec{q}}{q}) $. 
Similar for the graphs (c), (d) one needs the quantity 
$ J_i(\vec{q})=
\sum_{\vec{p}} G_{\pm}(p) G_{\mp}(p)(
\vec{e}_i 
\times \frac{\vec{q}}{q})=\tau (\vec{e}_i \times \frac{\vec{q}}{q}) $.  
One now has to discuss the different frequency combinations of (a), (b).
We make in the following the convention $ \Omega >0 $.
Because of the frequency constraints of the pole of the particle-hole channel
$ (1/\tau) \Gamma(\epsilon,\omega,q) $  one has 
some frequency constraint on the graphs to get this pole. 
Du to the different prefactor of the first terms 
$ J^{\pm}_1 $ of $ J^\pm $ one immediately gets for the nonvanishing  
$ (J^{\pm}_1)^2 $-
terms in $ \sigma^{D_{11},a}+ \sigma^{D_{11},b}$  
the two frequency combinations $\epsilon < 0, \epsilon+\Omega>0,
\epsilon+\omega +\Omega>0, \epsilon+\omega<0 $ and    
 $\epsilon < 0, \epsilon+\Omega<0,
\epsilon+\omega +\Omega>0, \epsilon+\omega>0 $. In each one of these 
two frequency combinations one gets 
$\sigma^{D_{11},a}+ \sigma^{D_{11},b}=2\sigma^{D_{11},a} $.
For the diagrams (a) (b) with reversed  Green's function direction one gets 
two similar frequency combinations of additive $ \sigma $ from (a), (b).    
Then one can carry out the $ \Omega $-integration \cite{al1} and 
gets for the conductivity  of the diagrams (a), (b) of  
current-combination $ (J_1^{\pm})^2 $ 
\begin{equation} \label{eq15}
 \delta \sigma_{ii}^{D_{11}}\approx-\frac{i 2 e^2}{2 \pi}
\int\limits_\Omega^{\frac{1}{\tau}}
 \frac{d \omega}{2\pi}\int \frac{d \vec{q}}{(2 \pi)^2}
\frac{\left(4J_{1,i}^+(q)J_{1,i}^+(q)\right)
\;\theta(Dq^2-|\omega|) }
{m \tau^2 \left[ D q^2-i\left|\omega+\Omega\right|\right]
 \left(-i |\omega| \gamma_q+\chi'_q q^2\right)} \;. 
\end{equation}
At finite temperature 
$ T \gg \Omega $ we can calculate the frequency-integral in 
$ \omega $ by using imaginary frequencies with an $ \omega $ integral 
cut-off at low frequencies $ \frac{1}{T} $. For $ V_q=(2\pi)
\frac{e^2}{\kappa} $ 
one gets 
\begin{equation} \label{eq16}
\delta \sigma_{ii}^{D_{11}}\approx\frac{2}{\pi h} 
\left(\log(k_F l)+i \frac{3}{8} \pi\right)  \log(T \tau) \;.
\end{equation}  
For $ V_q=(2\pi)\frac{e^2}{q} $ we make a partial fraction decomposition 
of the denominator of (\ref{eq15}) and get three additive terms
($ \Omega \to 0 $)
$ 1/2\, ( \pm 8 \pi i |\omega| \gamma_q +\sqrt{i|\omega|/D})\,
(i|\omega|/D)^{-\frac{3}{2}}\, (q\mp \sqrt{i|\omega|/D})^{-1} $ and 
$ i D/|\omega|\, (- 8 \pi i |\omega|\gamma_q +q)^{-1} $. Then one sees 
that the integral (\ref{eq15}) results in a finite value for $ T \to 0 $. 
Next we calculate $\sigma^{D_{11},a}+ \sigma^{D_{11},b}$ for the 
current combinations  $ (J^{\pm}_2)^2 $ and $ J^{\pm}_1 J^{\pm}_2 $. 
For example for the 
$ (J^{\pm}_2)^2 $ combination one gets for the conductivity to be nonvanishing
the two regimes  
$ \epsilon >0 $, $ \epsilon+\Omega<0 $ and $ \epsilon <0 $, 
$ \epsilon+\Omega>0 $. Integrating out the $ \Omega $-integral one 
gets two terms $ \Omega \int d\omega $, $ \int d\omega \,\omega $. The 
$ \log(T\tau) $-singularity  is cancelled for the 
$ \Omega \int d\omega $-terms of  
the two frequency ranges and 
the remaining two $ \int d\omega \,\omega $-integrals are infinite. These 
infinities are cancelled by similar infinities of the graphs $ (c), (d)$.  
Furthermore one can see with the help of 
$ J^{\pm} $, $ J $ that the $ \Omega \int d\omega $ integrals are cancel 
for every single graph $ (c) $, $ (d) $.             
In summary one can see, 
that the sum of the conductivities of the graphs (a), (b) for $ (J^{\pm}_2)^2$, $ J^{\pm}_1 J^{\pm}_2 $  
and of the graphs (c), (d) is zero for  
$ T \to 0$. 
One sees from the above analysis that the typical $\rho \rho $ 
part of the  CS-current is important to get a finite result for lower 
$ k_Fl $. \\  
We now compare our results with the experimental results of Rokhinson et al 
\cite{ro1}. They get  a $ \log (T) $ dependence for $ \sigma^{CS}_{xx} $ 
for temperature $ T < 500$mK . 
So we use $ V_q=2 \pi e^2/\kappa $ to compare our results with their 
measurements. 
With $ \sigma^{CS}_{xx}=
\sigma_0^{CS}+\lambda \frac{e^2}{h} \log(T \tau) $ they get 
$ \lambda_{ex}=(0.1,0.4,1.6) $ for Drude conductivities 
$ \sigma_0^{CS}=e^2/(2h) (k_F l)=(5.4 e^2/h, 16.0 e^2/h, 39.0 e^2/h) $.
From (\ref{eq14}), (\ref{eq16}) we get for the theoretical value 
$ \lambda_{th} $, $ \lambda_{th}=(1.81,2.53,3.1) $.
We see from these results that $ \lambda_{th} $ is getting better for 
 larger $ k_F l $.  \\
An improvement of the above calculation would be the calculation of 
the conductivity of diagrams which are
originating through the operations of section 2  from the 
corresponding Hartree self energy diagrams  of figure 2. 
For the $ D_{00} $-vertex this was done by Castellani et al. \cite{fi1}. 
They calculated higher order $ D_{00} $-vertex graphs  
and got for $ k_Fl  \gg 1 $, 
$ \sigma_{xx}^{CS} = e^2/(\pi h) (2-2\log(2)) \log(T \tau) $. This is 
only a small correction to (\ref{eq14}) and (\ref{eq16}) for $ k_F l \gg 1 $. 
Further one sees from $ D_{11}(0,q) \sim 1/q^2 $  
that the $ D_{11} $-Vertex can not give any contribution to 
$ \sigma^{CS}_{xx} $ in Hartree-diagrams \cite{al1}. \\
At least one could calculate the CS-conductivity from the diagrams of 
figure 2 for small deviations of the magnetic field $ B_{1/2} $. 
This means that the CFs are subject to an effective magnetic field 
 $ B^d=B-B_{1/2} $. For $ \omega^d_c\tau \ll 1 $, $ k_F l \gg1 $
one gets with the help of the formalism of Houghton et al. \cite{ho1},
$ J^\pm_{1,i}(\vec{q},B^d)=i \epsilon_F \tau^2(\mp
1/(1+(\omega_c^d \tau)^2)(\vec{e}_i \times \frac{\vec{q}}{q})
-(\omega^d_c \tau) 
(\vec{e_i}\cdot \frac{\vec{q}}{q})) $. 
Also one gets for 
the magnetic correction to the 
diffusion constant $ D $,   
$ D(B^d)=D/(1+(\omega_c^d \tau)^2) $ (this should also be considered 
in $\gamma_q$).   
In this $ k_F l \gg 1 $ limit, the only relevant graphs in figure (2) 
are (a), (b).
Because the second term in $ J^\pm_{1,i}(\vec{q},B^d) $ has the same 
behaviour as $ J_2^{\pm} $ for $ B^d=0 $ one sees from the 
above analysis that the conductivity 
$ \sigma^{a}_{xx}(B^d)+\sigma^{b}_{xx}(B^d) $
is infinite for 
$ |\omega^d_c \tau|>0 $. Since the graphs (c),(d) of figure 2 
give only lower order terms in $ k_F l $ this singularity can not be 
cancelled. So the next stage of improvement should be  
to consider Feynman-graphs which make $ \sigma_{xx}(B^d)$ finite for 
$ |\omega^d_c \tau|>0 $. This should also give better results for 
$ \sigma_{xx}(0)$. Further we have to remark that we don't
use a density of states correction at the fermi-level due to $ B_d \not=0 $
in the above formulas because the general statement of an infinite
conductivity is still correct.\\  
At least we have to remark that Khveshchenko \cite{kh1, kh2} calculated the 
conductivity of the graphs (a), (b) of figure 2 for $ k_F l \gg 1 $. He comes to 
other results than (\ref{eq16}) and also for the conductivity 
of $ V_q=2 \pi e^2/q $ in the large $ k_F l $ limit.
For example the difference to (\ref{eq16}) is a factor $ 1/4 $. 
This is exakt the result one gets if one assumes 
frequency constraints through vertex corrections $ \Gamma $
at the endpoints  of the $ D_{11} $ vertex as in the case of the 
$ D_{00}$ vertex \cite{al2}. These Feynman graphs 
have a vanishing conductivity 
 because of the current couplings at the endpoints of the $ D_{11} $ vertices. 
If one calculates the graphs (a), (b) without vertex corrections, 
one gets another result, as we have shown in this paper.

\end{document}